\documentclass[12pt]{iopart}
\usepackage{graphicx}
\usepackage{dcolumn}
\usepackage{color}
\usepackage{bm}
\usepackage{amssymb}
\usepackage{amsfonts}
\usepackage{epsf}
\usepackage{epsf}

\newcommand {\td} [2] {\frac{d #1}{d #2}}

\newcommand {\beq}{\begin{equation}}
\newcommand {\eeq}{\end{equation}}
\newcommand {\bea}{\begin{eqnarray}}
\newcommand {\eea}{\end{eqnarray}}

\begin{document}

\title {Parallel Coupling of Symmetric and Asymmetric Exclusion Processes}
\author{K.Tsekouras\dag  and  A.B.Kolomeisky\ddag}
\address{\dag Department of Physics, Rice University, Houston, TX 77005-1892}
\address{\ddag Department of Chemistry, Rice University, Houston, TX 77005-1892}

\begin{abstract}\\
A system consisting of two parallel coupled channels where particles in one of them follow the rules of totally asymmetric exclusion processes (TASEP) and in another one move as in symmetric simple exclusion processes (SSEP) is investigated theoretically. Particles interact with each other via hard-core exclusion potential, and in the asymmetric channel they can only hop in one direction, while on the symmetric lattice particles jump in both directions with equal probabilities. Inter-channel transitions  are also allowed at every site of both lattices. Stationary state properties of the system are solved exactly in the limit of strong couplings between the channels. It is shown that strong symmetric couplings between totally asymmetric and symmetric channels lead to an effective partially asymmetric simple exclusion process (PASEP) and properties of both channels become almost identical. However, strong asymmetric couplings between symmetric and asymmetric channels yield an effective TASEP with nonzero particle flux in the asymmetric channel and zero flux on the symmetric lattice. For intermediate strength of couplings between the lattices a vertical cluster mean-field method is developed. This approximate approach treats exactly particle dynamics during the vertical transitions between the channels and it neglects the correlations along the channels. Our calculations show that in all cases there are three stationary phases defined by particle dynamics at entrances, at exits or in the bulk of the system, while phase boundaries  depend on the strength and symmetry of couplings between the channels.  Extensive Monte Carlo computer simulations strongly support our theoretical predictions. Theoretical calculations and computer simulations predict that inter-channel couplings have a strong effect on stationary properties. It is also argued that our results might be relevant for understanding  multi-particle dynamics of motor proteins. \end{abstract}


\ead{tolya@rice.edu}

\submitto{\JPA}

\maketitle

\section{Introduction}

The majority of systems in nature operate far from equilibrium, but there is no developed theoretical framework for comprehensive analysis of non-equilibrium processes. In this situation, a critical role for understanding different complex phenomena in Chemistry, Physics and Biology is played by a class of  low-dimensional non-equilibrium  multi-particle models known as asymmetric simple exclusion processes (ASEP) \cite{derrida98,schutz,schutz03,blythe07,derrida07}. ASEP are stochastic models where particle interact via an exclusion potential and move along discrete lattices. Mechanisms of many non-equilibrium processes, such as biological transport, kinetics of protein synthesis and biopolymerization, car traffic, and hopping of quantum dots, have become better understood due to the successful description via asymmetric exclusion models \cite{macdonald68,shaw03,shaw04,chou04,chou99,klumpp03,mirin03,klumpp05,parmeggiani03,nishinari05,evans07,helbing01,chowdhury00}.

Most theoretical studies of exclusion processes concentrate on single-lane systems where important exact solutions have been obtained in several cases \cite{derrida98,blythe07}. Recently, a lot of attention has also been devoted to parallel multi-chain exclusion processes \cite{popkov01,pronina04,pronina05,mitsudo05,harris05,pronina06,reichenbach06,pronina07,reichenbach07,juhasz07,tsekouras08,jiang08}. The study of these models has been greatly stimulated by  experimental advances in analysis of motor proteins dynamics along cytoskeleton filaments, in transport of mesoscopic quantum systems and in vehicular traffic processes \cite{helbing01,reichenbach07,howard}. In parallel multi-chain exclusion processes particles can jump along the horizontal chains, but they can also switch stochastically between different lanes. Theoretical analysis of different multi-chain ASEP suggests that coupling between the channels strongly influences stationary-state phases and particle  properties. It can produce a complex dynamic behavior, leading to many unusual phenomena, such localized domain walls and symmetry breaking \cite{pronina06,reichenbach07,juhasz07,jiang08}.

Investigations of multi-chain exclusion processes mostly involves coupling of asymmetric channels where the direction of particles motion is biased at each site on all channels. The aim of the present paper is to analyze parallel coupling of symmetric and asymmetric exclusion processes. The problem is motivated by the cellular transport of motor proteins along rigid protein filaments such as actin filaments or microtubules \cite{howard}. Motor protein molecules can move mostly in one direction when they are tightly bound to protein lattices. Occasionally, motor proteins might dissociate from the filament to the surrounding solution where they perform unbiased diffusional motion. Freely diffusing molecules can also bind to  protein filaments. There is only one previous theoretical work that investigates coupling of TASEP and SSEP, although  periodic boundary conditions and symmetric transition rates between the lanes are assumed \cite{dickman07}. Using several mean-field approaches and extensive computer simulations it was found that there is unequal redistribution of particles between different channels depending on the densities  \cite{dickman07}. In a related study, Lipowsky and coworkers investigated transport of molecular motors in open tube that contains a single filament \cite{klumpp03,klumpp05}. Bound to the filament particles undergo asymmetric exclusion process, while the unbound molecular motors diffuse freely in the tube around the linear chain. The analysis of the molecular motor transport via tube-like compartments, performed with the help of mean-field methods and Monte Carlo simulations, revealed that there are three stationary phases with phase transitions specified by the precise choice of boundary conditions  \cite{klumpp03}. In our work we investigate a two-channel system consisting asymmetric and symmetric exclusion lanes with open boundary conditions, and with symmetric and asymmetric transition rates between the channels.

This paper is organized as follows. In section 2  a detailed description of the model is given, and exact solutions for strong couplings and approximate solutions for intermediate couplings are presented. In section 3, we discuss Monte Carlo computer simulations and compare them with theoretical predictions. The final section 4 summarizes and concludes.

\section{Theoretical analysis}

\subsection{Model}

Our model  consists of two parallel one-dimensional lattices as illustrated in figure 1. Both lanes have $L$ sites, and we are interested in obtaining thermodynamic limit results when $L \gg 1$.  Particles move along the channels by hopping between the lattice sites with the exclusion potential, i.e., each site can be occupied by no more than one particle. In our model we apply a random sequential update when at each time step dynamics at randomly chosen site is followed. Particles can enter the system with the rate $0 < \alpha \le 1$ if one or both first sites on channel 1 and 2 are not occupied. Similarly, particles exit the system with the rate $0 < \beta \le 1$ if any of last sites are occupied. In the bulk of the system  dynamic rules depend on the lattice: see figure 1. The particle at site $i$ can switch to the same site on the lattice 2 with the rate $w_{1}$ if this site is empty, or with the rate $1-w_{1}$ it hops to  unoccupied site $i+1$ on the lattice 1. However, if the site $i$ in the second channel is occupied, the particle jumps in the horizontal direction to the right with the rate 1 if the forward site is available.  The particle at the site $i$ on  the lattice 2 can move vertically with the rate $w_{2}$ if the upper site $i$ is free, or it can jump horizontally in either direction with the rate $(1-w_{2})/2$ if sites $i+1$ or $i-1$ are available. However, if the upper site $i$ is already occupied, the particle can move with the rate 1/2 in the forward or backward directions, assuming that any of these moves are not blocked by already present particles at sites $i-1$ or $i+1$. The total transition rate out of every site $i$ in any channel is equal to one. When the transition rates between the channels are equal ($w_{1}=w_{2}$) the coupling is symmetric, while for $w_{1} \ne w_{2}$ the coupling between the lattices is asymmetric.

\begin{figure}[h]
\centering \includegraphics[scale=0.4,clip=true]{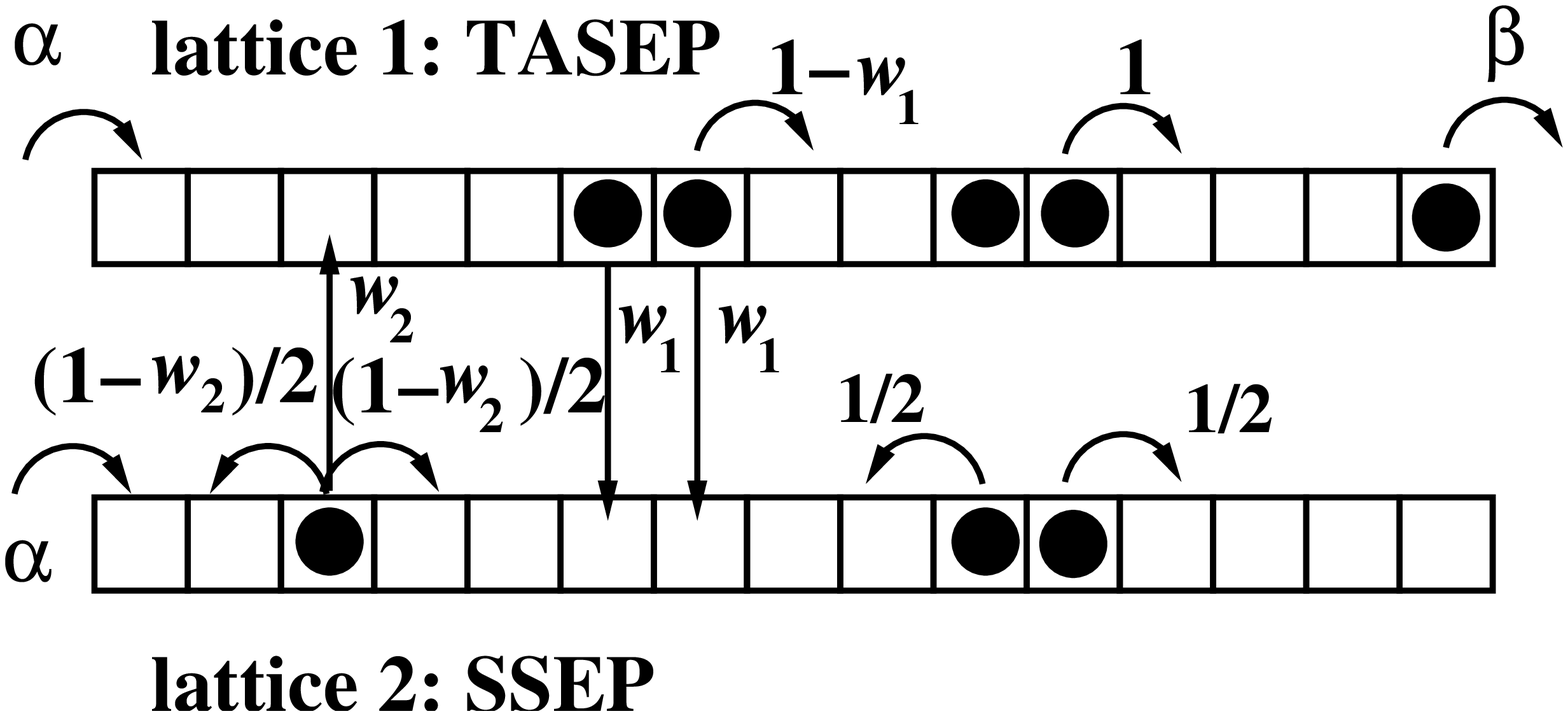} \caption{Schematic picture of two-channel system that couples symmetric and asymmetric exclusion processes.  On the lattice 1 (upper) particles can move only to the right, while on the lattice 2 (lower) there is no preference in the direction of motion. The inter-channel transitions rates are $w_{1}$ and $w_{2}$. Allowed transitions are shown by arrows. Entrance rates at both lattices are equal to $\alpha$ and exit rates are equal to $\beta$.}
\end{figure}

When the inter-channel transition rates are equal to zero ($w_{1}=w_{2}=0$), the system decouples into two independent single-lane exclusion processes: the upper channel becomes a totally asymmetric process and the lower channel is a symmetric process. Exact solutions for single-lane TASEP and SSEP are known \cite{derrida98,derrida07}, and they provide a full description of all dynamic properties at large times. For TASEP there are three stationary-state phases. If the entrance to the lattice is a rate-limiting step, which happens for $\alpha < \beta$ and $\alpha < 1/2$, the system is found in a low-density (LD) phase with  the particle current and  bulk density  given by 
\begin{equation}\label{ld1}
J_{LD}=\alpha(1-\alpha), \quad \rho_{bulk,LD}=\alpha.
\end{equation}
However, when exit of particles  controls the overall dynamics ($\beta < \alpha$ and $\beta < 1/2$), the system is in a high-density  (HD) phase where the stationary current and bulk density are
\begin{equation}\label{hd1}
J_{HD}=\beta(1-\beta), \quad \rho_{bulk,HD}=1-\beta.
\end{equation}
In the third phase, called a maximal-current (MC), the dynamics is governed by bulk processes ($\alpha > 1/2$ and $\beta > 1/2$), and stationary properties are the following,
\begin{equation}\label{mc1}
J_{MC}=\frac{1}{4}, \quad \rho_{bulk,MC}=\frac{1}{2}.
\end{equation}

The stationary properties of SSEP are much simpler with only one non-equilibrium phase at all conditions \cite{derrida07}. For the lattice with $L$ sites the density profile is linear, 
\begin{equation}
\rho=\frac{L-i+1/\beta}{L+1/\alpha+1/\beta-1}.
\end{equation}
The average current in the steady-state of SSEP is given by
\begin{equation}
J=\frac{1}{L+1/\alpha+1/\beta-1}.
\end{equation}
It can be easily seen that in the limit of $L \rightarrow \infty$ the current in the SSEP is approaching zero.

There are very few exact results for multi-channel exclusion processes obtained by mapping them into effective single-lane exclusion models \cite{pronina04}. Typical approaches to analyze coupled multi-channel systems involve various mean-field treatments supported by computer simulations \cite{popkov01,pronina04,pronina05,mitsudo05,harris05,pronina06,reichenbach06,pronina07,reichenbach07,juhasz07,tsekouras08,jiang08}. Probably, one of the most successful approximate approaches is a vertical-cluster mean-field method \cite{pronina04,pronina06,pronina07,tsekouras08}, that describes the dynamics of inter-channel transitions explicitly and neglects correlations for horizontal transitions. We will also utilize this method for analyzing parallel coupling between TASEP and SSEP.  Each vertical cluster can be described by introducing functions $P_{ij}$ ($i,j=0,1$) that define the probability of different states. $P_{00}$ corresponds to the state when both sites of the vertical cluster are empty, $P_{11}$ describes the state with both sites occupied, and $P_{10}$ and $P_{01}$ specify partially filled vertical clusters with the occupied site on the channel 1 or 2, respectively. These probability functions are related via a normalization condition,  
\begin{equation}\label{norm}
P_{00}+P_{10}+P_{01}+P_{11}=1.
\end{equation}

\subsection{Strong couplings: Exact results}

In order to understand mechanisms of coupling between TASEP and SSEP, it is instructive to consider strong coupling regimes. First, let us analyze the case of symmetric coupling with $w_{1}=w_{2}=0$. In this case in the stationary-state limit it is not possible to observe the vertical configuration $\{00\}$ with two empty sites. This is because this configuration can only be obtained by moving the particle horizontally if the previous state of the same cluster was $\{10\}$ or $\{01\}$. However, the rates for these transitions are zero ($1-w_{1}=1-w_{2}=0$), and we conclude that for any bulk vertical cluster $P_{00}=0$. It can be also argued that symmetry of the coupling requires to have $P_{01}=P_{01}$ for any bulk vertical cluster. Then in the system there are only two types of vertical clusters: half-filled and fully filled. One can view $\{11\}$ vertical clusters as  effective ``particles'' and half-filled vertical clusters as effective ``holes.''  These ``particles'' can advance forward with the rate $p=3/4$, or they can move backward with the rate $q=1/4$.  They enter the system with the effective rate $\alpha_{eff}=\alpha$ and leave it with the effective rate $\beta_{eff}=2\beta$. The factor 2 comes from the fact that in the last vertical cluster both particles can exit independently.  Thus the two-channel system is mapped into a new effective single-channel partially asymmetric simple exclusion process (PASEP), for which {\it exact} solutions are known \cite{sandow94}.    

Similarly to totally asymmetric exclusion processes there are three stationary phases in PASEP \cite{sandow94}. The LD phase exists for $\alpha < \beta$ and $\alpha < (p-q)/2$, and the stationary properties of this phase can be written as
\begin{equation}\label{ld2}
J_{LD}=\frac{\alpha(p-q-\alpha)}{p-q}, \quad \rho_{bulk,LD}=\frac{\alpha}{p-q}.
\end{equation}
The HD phase can be found for $\beta < \alpha$ and $\beta < (p-q)/2$ with the particle current and bulk density given by
\begin{equation}\label{hd2}
J_{HD}=\frac{\beta(p-q-\beta)}{p-q}, \quad \rho_{bulk,HD}=1-\frac{\beta}{p-q}.
\end{equation}
In the MC phase ($\alpha > (p-q)/2$ and $\beta > (p-q)/2$) the stationary properties are the following,
\begin{equation}\label{mc2}
J_{MC}=\frac{p-q}{4}, \quad \rho_{bulk,MC}=\frac{1}{2}.
\end{equation}
Applying these results to our system, it is trivial to show that the LD phase is specified by $\alpha < 2 \beta$ and $\alpha < 1/4$ with the particle current and bulk density given by
\begin{equation}\label{ld3}
J_{LD}=\alpha(1-2 \alpha), \quad \rho_{bulk,LD}=1/2+\alpha.
\end{equation}
The HD phase exists  for $\alpha > 2 \beta$ and $\beta < 1/8$ with the following stationary properties,
\begin{equation}\label{hd3}
J_{HD}=2 \beta( 1-4 \beta), \quad \rho_{bulk,HD}=1- 2 \beta.
\end{equation}
For $\alpha > 1/4$ and $\beta > 1/8$ we have the MC phase with
\begin{equation}\label{mc3}
J_{MC}=1/8, \quad \rho_{bulk,MC}=\frac{1}{2}.
\end{equation}
The resulting density profiles for strong coupling limit are shown in figures 2a, 3a and 4a, while the phase diagram is outlined in figure 5a.

Exact solutions via mapping to a single-channel exclusion process can also be found for strong asymmetric couplings. Let us show this for the case of $w_{1}=1$ and $w_{2}=0$. For the arbitrary bulk vertical cluster at site $i$ we have  $P_{10}=0$, because there should not be the overall vertical current in the system at large times. It can be argued that the whole lattice 2 is fully occupied at large times after the system reaches a steady state. If any vacancy appears then it will be quickly filled by vertical transition form the upper channel. This observation suggests that bulk vertical clusters can only be found in two states, $\{11\}$ and $\{01\}$ with $P_{00}=P_{10}=0$.  Note, however, that other cluster states might exist near the boundaries, but they should not affect the overall dynamics in the system. We associate the fully filled vertical clusters $\{11\}$ with  ``particles'', while the vertical clusters $\{01\}$ can be viewed as ``holes.'' There is the particle flux in the upper channel, and there is no current on the lattice 2 and the bulk density in the channel 2 is always equal to one. Thus, we mapped the system that couples asymmetric and symmetric exclusion process into  new effective TASEP in the limit of strong asymmetric coupling. 

The effective entrance rate for the ``particles'' is equal to $\alpha_{eff}=\alpha$. However, the exit process should be considered more carefully. The overall particle flux to leave the channels can be written as
\begin{equation}
J_{exit}=\beta_{eff}P_{11}=\beta (2 P_{11}+P_{01})=\beta (1+P_{11}),
\end{equation}
and it should be equal to the bulk current, 
\begin{equation}
J_{bulk}=P_{11}(1-P_{11}). 
\end{equation}
It leads to the following relation for density of fully filled vertical clusters at the end of the system,
\begin{equation}
P_{11}=\frac{1-\beta+\sqrt{\beta^{2}-6\beta+1}}{2},
\end{equation}
which also yields the effective exit rate,
\begin{equation}
\beta_{eff}=\frac{1+\beta-\sqrt{\beta^{2}-6\beta+1}}{2}.
\end{equation}
Now, using known results for TASEP we can predict that in the two-channel exclusion system with strong asymmetric coupling there are three stationary phases. Entrance dominated low-density phase exists for $\alpha <1/2$ and $\beta > \frac{\alpha(1-\alpha)}{(2-\alpha)}$. Here the stationary properties are
\begin{equation}\label{ld4}
J_{LD}^{(1)}=\alpha(1-\alpha), \quad \rho_{bulk,LD}^{(1)}=\alpha.
\end{equation}
Exit dominated high-density phase is found for $\beta <1/6$ and $\beta < \frac{\alpha(1-\alpha)}{(2-\alpha)}$ with the following particle current and bulk density on the lattice 1:
\begin{equation}\label{hd4}
J_{HD}^{(1)}=\frac{\beta( 3- \beta + \sqrt{\beta^{2}-6\beta+1})}{2}, \quad \rho_{bulk,HD}^{(1)}=\frac{1-\beta+\sqrt{\beta^{2}-6\beta+1}}{2}.
\end{equation}
Finally, in the maximal-current phase (for $\alpha > 1/2$ and $\beta > 1/6$) we have
\begin{equation}\label{mc4}
J_{MC}^{(1)}=1/4, \quad \rho_{bulk,MC}^{(1)}=\frac{1}{2}.
\end{equation}
Density profiles for strong asymmetric coupling  with $w_{1}=1$ and $w_{2}=0$ are shown in figures 2c, 3c and 4c, and the phase diagram is presented in figure 5c.

The other case of the asymmetric coupling, when $w_{1}=0$ and $w_{2}=1$, can be easily analyzed if we recall the particle-hole symmetry of the system. The flux of particles moving from the left to right can be viewed as a flux of holes moving in opposite direction. Then stationary properties in this case can be obtained from Eqs. (\ref{ld4}), (\ref{hd4}) and (\ref{mc4}) derived above if we perform the symmetry operations $\alpha \leftrightarrow \beta$ and $0 \leftrightarrow 1$. In this case, the nonzero particle current will be found only in the upper (asymmetric) channel, but there will be no particles  and no flux in the bulk of the symmetric lattice.

\subsection{Intermediate couplings: Approximate theory}

When couplings between asymmetric and symmetric exclusion processes are not strong ($w_{1} <1$ and/or $w_{2} < 1$) it is not possible to solve the system exactly via mapping procedure. Then an approximate theory should be developed. We will use the vertical-cluster mean-field approach \cite{pronina04,pronina06,tsekouras08} that was successful in description of other two-channel exclusion processes. 

The overall properties of the system can be found by monitoring changes in four vertical clusters at each site. Assuming that the behavior is uniform along the lattices, the dynamics of bulk vertical clusters is governed by three independent master equations:
\begin{equation}\label{me1}
\td{P_{11}}{t}=(2-w_1-w_2) P_{10} P_{01}-2P_{11} P_{00},
\end{equation}
\begin{equation}\label{me2}
\td{P_{10}}{t}=w_2 P_{01}-w_1 P_{10}+2P_{11} P_{00}-(2-w_1-w_2) P_{10} P_{01},
\end{equation}
\begin{equation}\label{me3}
\td{P_{01}}{t}=w_1 P_{10}-w_2 P_{01}+2P_{11} P_{00}-(2-w_1-w_2)P_{01} P_{10}.
\end{equation}
At large times the system reaches steady state, implying that $\td{P_{ij}}{t}=0$ for $i,j=0,1$. Then from equations (\ref{me2}) and (\ref{me3}) we can immediately conclude that
\begin{equation}\label{vert_eq}
w_2 P_{01}=w_1 P_{10}.
\end{equation}
This expression can be understood as an equilibrium for vertical transitions between the channels in the bulk. Then substituting this relation along with the normalization condition (\ref{norm}) into equation (\ref{me1}) produces
\begin{equation}\label{relation}
(2-w_1-w_2)\frac{w_1}{w_2} P_{10}^2+2(1+\frac{w_1}{w_2})P_{11} P_{10}-2P_{11}(1-P_{11})=0.
\end{equation}
We expect that, similarly to the cases of strong couplings, there are three stationary phases. To obtain stationary properties explicitly we need expressions for entrance, exit and bulk currents through the system,
\begin{equation}
J_{entrance}=\alpha(2P_{00}+P_{10}+P_{01}),
\end{equation}
\begin{equation}
J_{bulk}=\left[ P_{11}+(1-w_1)P_{10} \right ](P_{00}+P_{01}),
\end{equation}
\begin{equation}
J_{exit}=\beta \left[ 2 P_{11}+(1-w_1)P_{10}+(1-w_2)P_{01} \right].
\end{equation}
Substituting into these relations the values for $P_{01}$ and $P_{00}$ from the  equilibrium for switching between the channels (\ref{vert_eq}) and from the normalization (\ref{norm}) we obtain
\begin{equation}
J_{entrance}=\alpha \left [ 2(1-P_{11}) - (1 + \frac{w_{1}}{w_{2}})P_{10} \right ],
\end{equation}
\begin{equation}\label{current_bulk}
J_{bulk}=\left[ P_{11}+(1-w_1)P_{10} \right ] (1-P_{11}-P_{10}),
\end{equation}
\begin{equation}
J_{exit}=\beta \left[ 2 P_{11}+(1-2w_1 + \frac{w_{1}}{w_{2}})P_{10} \right].
\end{equation}

The general strategy for solving the system is the following. From equation (\ref{relation}) we express $P_{10}$ in terms of $P_{11}$ and then all stationary quantities will depend only on one variable. The conditions for existence and dynamic properties of entrance-dominated LD phase and exit-dominated HD phase can be found from the condition of the stationarity of the current. The MC phase can be determined by solving $\frac{\partial J_{bulk}}{\partial P_{11}}=0$. The bulk densities in each channel can be calculated from
\begin{equation}
\rho_{bulk}^{(1)}=P_{11}+P_{10}, \quad \rho_{bulk}^{(2)}=P_{11}+P_{01}.
\end{equation} 

We now proceed to analyze symmetric couplings with $w_{1}=w_{2}=w<1$. The equilibrium for vertical transitions gives us $P_{01}=P_{10}$ and equation (\ref{relation}) simplifies into
\begin{equation}
(1-w)P_{10}^2+2P_{11} P_{10}-P_{11}(1-P_{11})=0,
\end{equation}
which yields the following solution,
\begin{equation}
P_{10}=\frac{\sqrt{P_{11}[1-w + w P_{11})]}-P_{11}}{1-w}.
\end{equation}
The properties of LD phase can be computed from the condition $J_{entrance}=J_{bulk}$ which leads to
\begin{equation}\label{P11LD}
P_{11}=\frac{\sqrt{(1-w)^2+16w\alpha^2}-(1-w)}{2w}.
\end{equation}
The expression for the current can be written in terms of $P_{11}$ as
\begin{equation}\label{currLD}
J_{LD}=2\alpha \ \frac{ \left [ 1-w+w P_{11} - \sqrt{P_{11}(1-w+w P_{11})} \right ]}{1-w},
\end{equation}
while the bulk densities in both channels are equal to each other, and they are given by
\begin{equation}\label{bulk_dens}
\rho_{bulk}^{(1)}=\rho_{bulk}^{(2)}=\frac{\sqrt{P_{11}[1-w + w P_{11})]}-w P_{11}}{1-w}.
\end{equation}
Substituting equation (\ref{P11LD}) into equations (\ref{currLD}) and (\ref{bulk_dens}) we obtain
\begin{equation}
J_{LD}=\frac{\alpha}{1-w} \left [ \sqrt{(1-w)^2+16w\alpha^2} +(1-w) -4 \alpha  \right ],
\end{equation}
and
\begin{equation}
\rho_{bulk}^{(1)}=\rho_{bulk}^{(2)}= \frac{\left [ (1-w) + 4 \alpha - \sqrt{(1-w)^2+16w\alpha^2} \right ] }{2(1-w)}.
\end{equation}

Similarly, the properties of HD phase follow from the relation $J_{exit}=J_{bulk}$, from which we obtain
\begin{equation}\label{P11HD}
P_{11}=\frac{2w(1-2\beta)-1+\sqrt{1-8w\beta +16 w\beta^{2}}}{2w}.
\end{equation}
The stationary current can be written as
\begin{equation}
J_{HD}=2 \beta \sqrt{P_{11}(1-w+w P_{11})},
\end{equation}
and the bulk densities are are the same as in equation (\ref{bulk_dens}). Using equation (\ref{P11HD}) the explicit expressions for the current and densities are the following:
\begin{equation}
J_{HD}= \beta \left( 1-4 \beta + \sqrt{1-8w\beta +16 w\beta^{2}} \right ),
\end{equation}
\begin{equation}
\rho_{bulk}^{(1)}=\rho_{bulk}^{(2)}= 1-2 \beta.
\end{equation}
The surprising result is that the bulk densities in HD phase, in contrast to the LD phase, are independent of the coupling strength.

In the MC phase the current can be derived from  equation (\ref{current_bulk}),
\begin{equation}
J_{MC}= \sqrt{P_{11}(1-w+w P_{11})} \ \frac{\left[ 1-w+w P_{11} - \sqrt{P_{11}(1-w+w P_{11})} \right ]}{1-w}.
\end{equation}
From the condition of maximum of the current, $\frac{\partial J_{bulk}}{\partial P_{11}}=0 $, we can obtain the explicit form of $P_{11}$ for every value of the inter-channel transition rate $w$,
\begin{equation}
P_{11}=\frac{3w-2}{6w}+\frac{64 -144 w+144w^2}{192w \sqrt[3]{Y}}+\frac{\sqrt[3]{Y}}{12w},
\end{equation}
where 
\begin{equation}
Y=8-27w^2+27w^3+3\sqrt{3}\sqrt{16w-68w^2+115w^3-90w^{4}+27w^{5}}.
\end{equation}
Boundaries between stationary phases can be determined from the particle currents for each regime at transition lines. For example, HD and LD phase are separated by a curve given by
\begin{equation}
 \frac{\alpha \left( 1-w - 4 \alpha + \sqrt{(1-w)^{2}+16 w \alpha^{2}} \right) }{1-w}= \beta \left( 1-4 \beta + \sqrt{1-8w\beta +16 w\beta^{2}} \right).
\end{equation} 
Note, that when $w=1$ we recover the results for strong symmetric coupling obtained in section 2.2, as expected.

For general case of intermediate couplings between symmetric and asymmetric exclusion channels ($w_{1} \ne w_{2}$) we utilize the same approach. The solution of equation (\ref{relation}) gives us
\begin{equation}
P_{10}=\frac{-P_{11}(w_1+w_2)+\sqrt{(w_1+w_2)^2P_{11}^2+2w_2P_{11}(2w_1-w_1w_2-w_1^2)(1-P_{11})}}{w_1(2-w_2-w_1)}.
\end{equation}
Then this equation can be used to express all stationary properties in terms of only one variable, $P_{11}$, and explicit calculations can be done as described above for the symmetric coupling.

\begin{figure}[h]
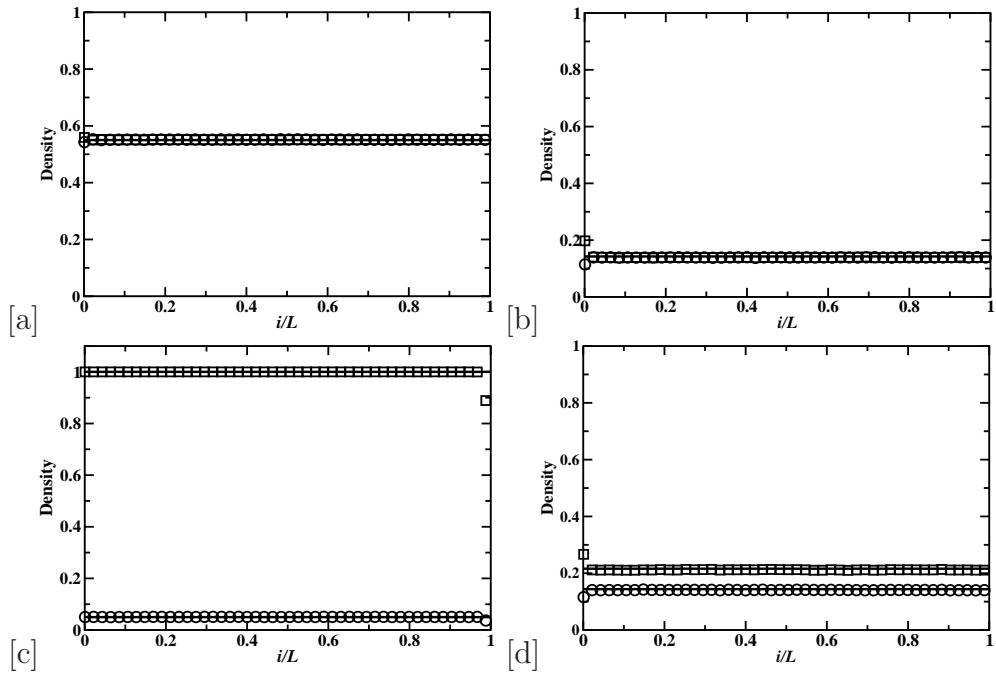

\centering [a]\includegraphics[scale=0.25,clip=true]{Fig2a.eps}
           [b]\includegraphics[scale=0.25,clip=true]{Fig2b.eps}
           [c]\includegraphics[scale=0.25,clip=true]{Fig2c.eps}
           [d]\includegraphics[scale=0.25,clip=true]{Fig2d.eps}
\caption{Density profiles for the entrance-dominated (LD) phase in a two-channel system that couples TASEP and SSEP for $\alpha = 0.05$ and  $\beta=0.75$. a)  $w_1=w_2=1$; b)  $w_1=w_2=1/3$; c)  $w_1=1$ and $w_2=0$; and d)  $w_1=0.4$ and $w_2=0.25$. Lines are theoretical predictions for bulk densities, while symbols correspond to Monte Carlo simulations: circles represent the upper (asymmetric) channel, squares are for the lower (symmetric) channel.}
\end{figure}

\begin{figure}[h]
\centering [a]\includegraphics[scale=0.25,clip=true]{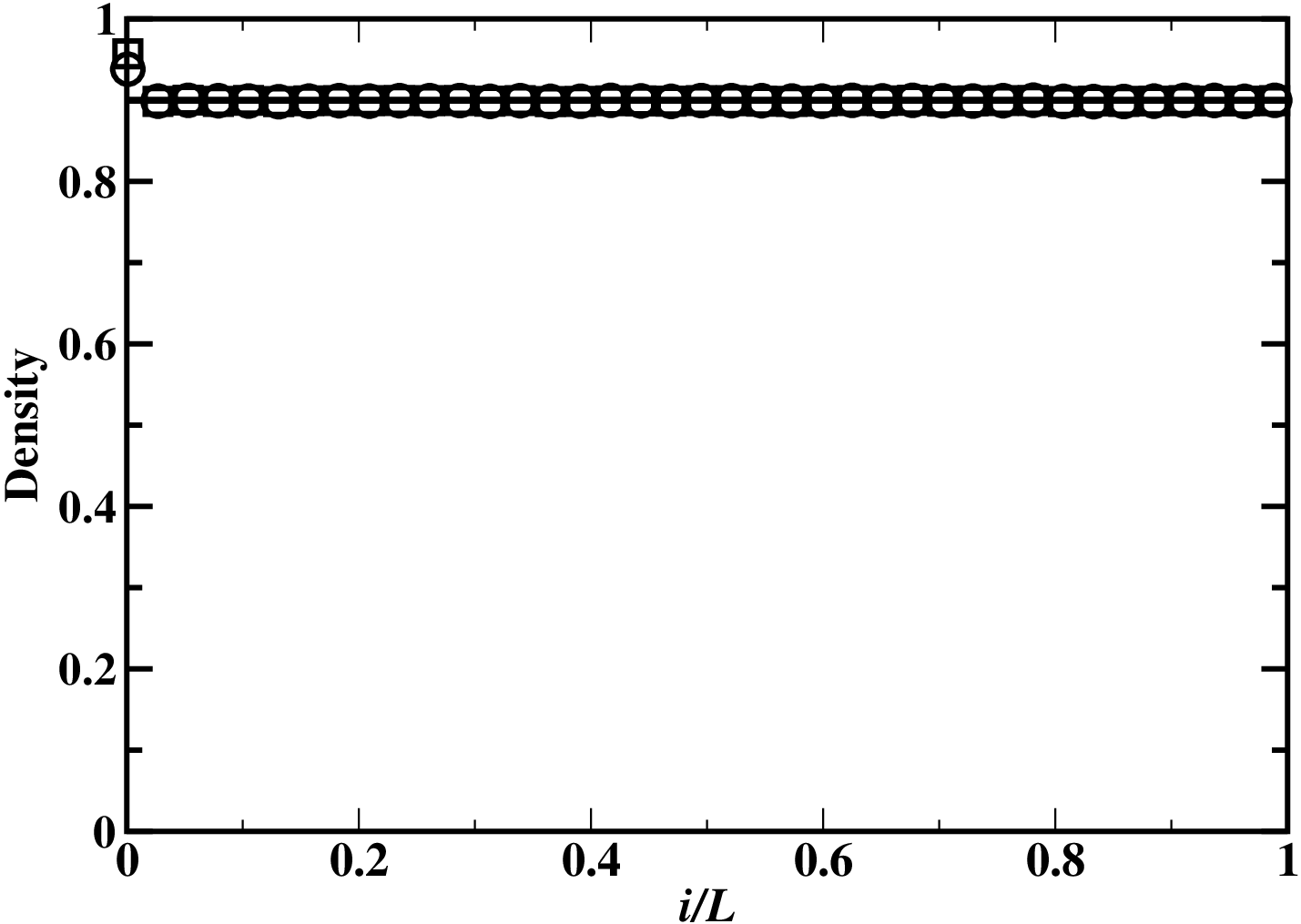}
           [b]\includegraphics[scale=0.25,clip=true]{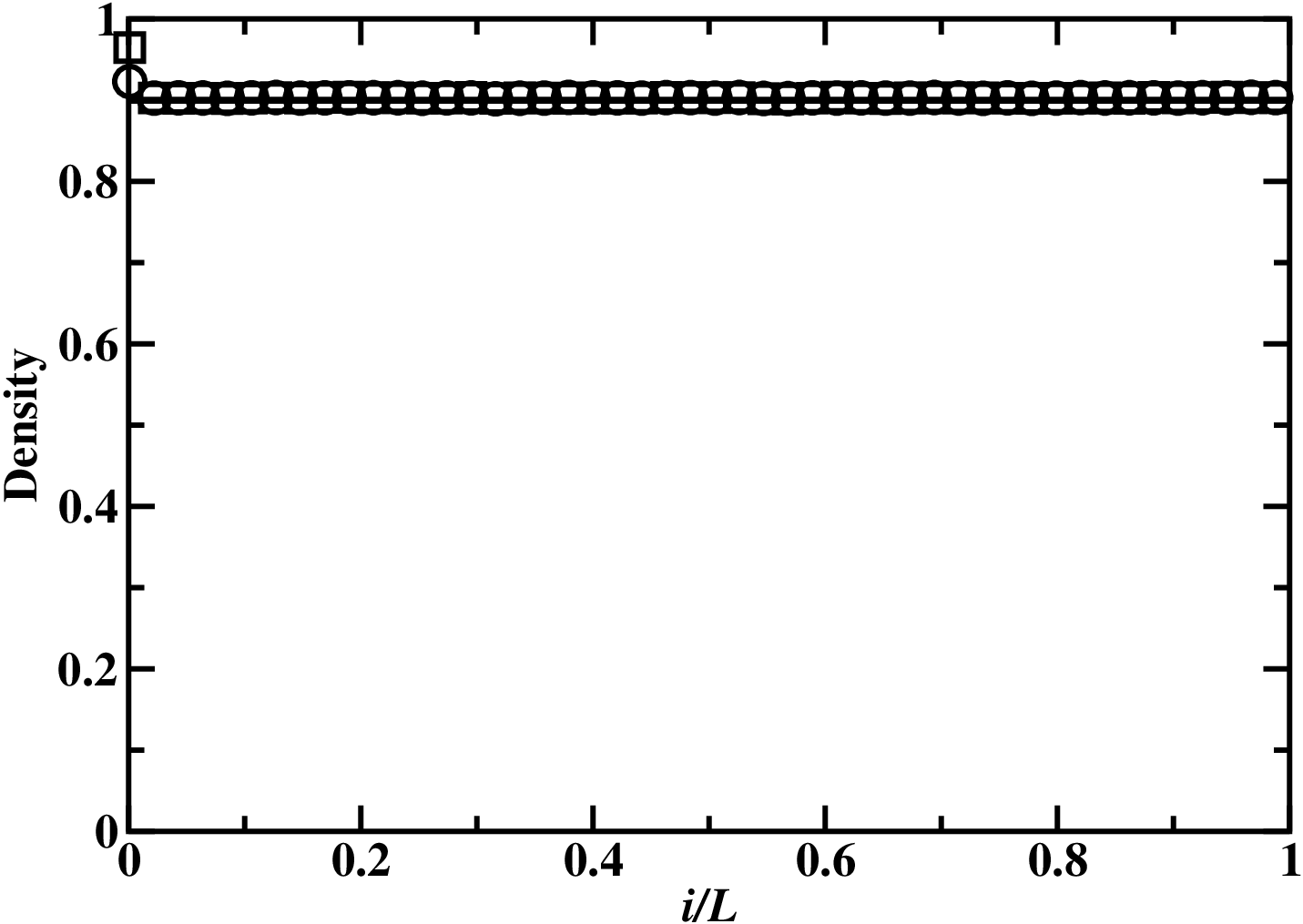}
           [c]\includegraphics[scale=0.25,clip=true]{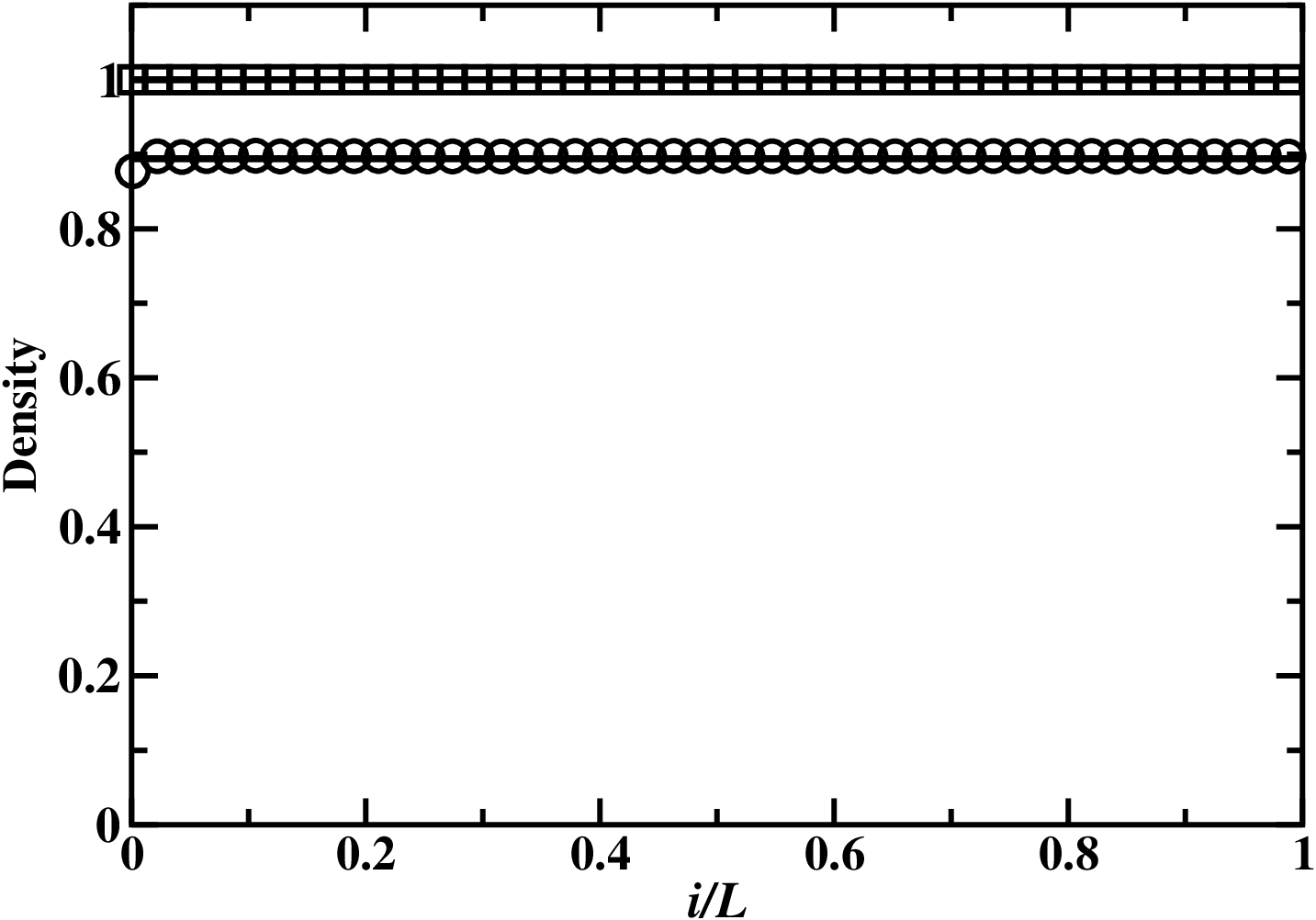}
           [d]\includegraphics[scale=0.25,clip=true]{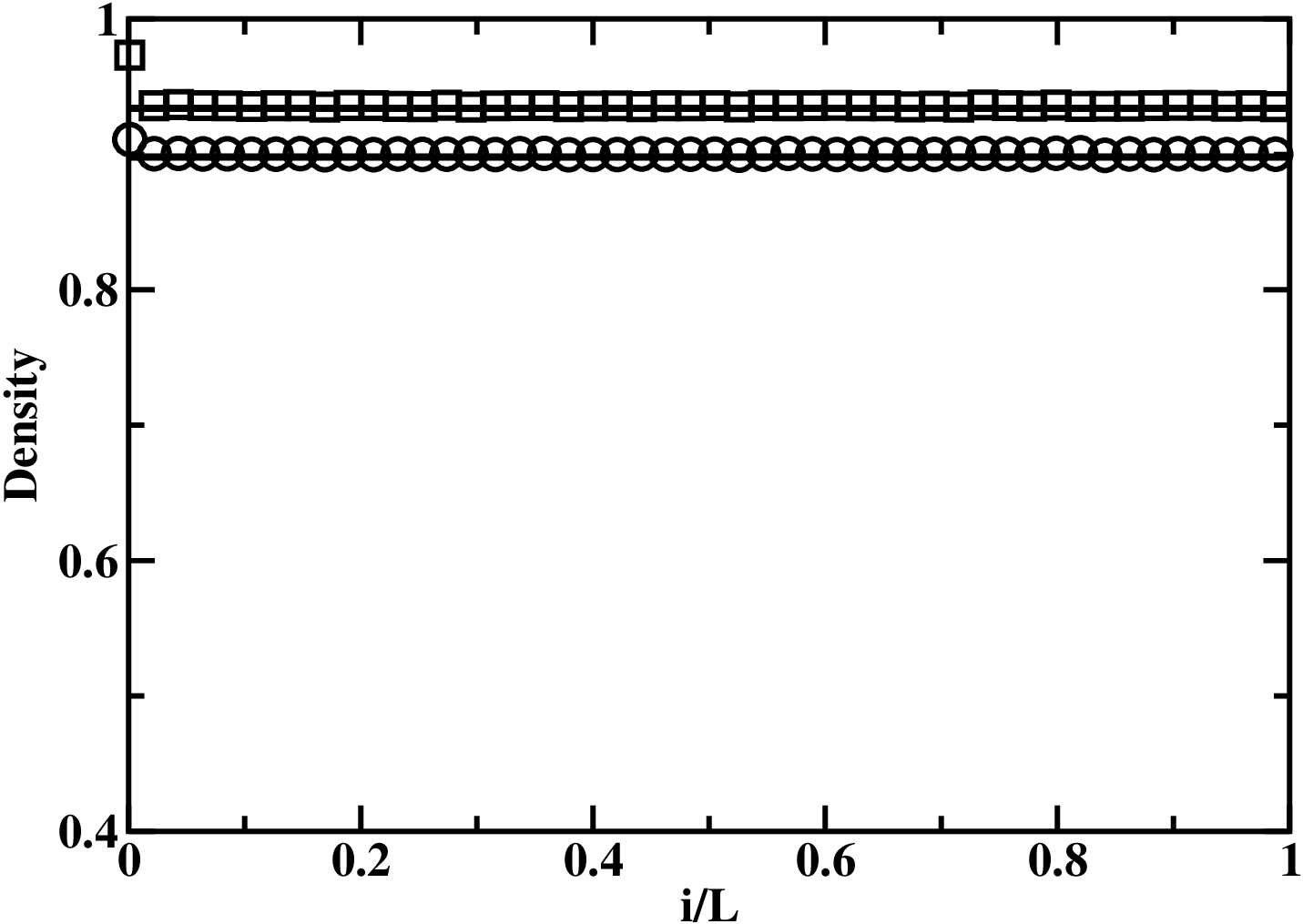}
\caption{Density profiles for the exit-dominated (HD) phase in a two-channel system that couples TASEP and SSEP for $\alpha = 0.75$ and   $\beta=0.05$: a)  $w_1=w_2=1$; b) $w_1=w_2=1/3$; c)  $w_1=1$ and $w_2=0$; and d) $w_1=0.4$ and $w_2=0.25$. Lines are theoretical predictions for bulk densities, while symbols correspond to Monte Carlo simulations: circles represent the upper (asymmetric) channel, squares are for the lower (symmetric) channel.}
\end{figure}

\begin{figure}[h]
\centering [a]\includegraphics[scale=0.25,clip=true]{Fig4a.eps}
           [b]\includegraphics[scale=0.25,clip=true]{Fig4b.eps}
           [c]\includegraphics[scale=0.25,clip=true]{Fig4c.eps}
           [d]\includegraphics[scale=0.25,clip=true]{Fig4d.eps}
\caption{Density profiles for the maximum current (MC) phase in a two-channel system that couples TASEP and SSEP for $\alpha = \beta=0.75$: a)  $w_1=w_2=1$; b)  $w_1=w_2=1/3$; c)  $w_1=1$ and $w_2=0$; and d) $w_1=0.4$ and $w_2=0.25$. Lines are theoretical predictions for bulk densities, while symbols correspond to Monte Carlo simulations: circles represent the upper (asymmetric) channel, squares are for the lower (symmetric) channel.}
\end{figure}

\begin{figure}[h]
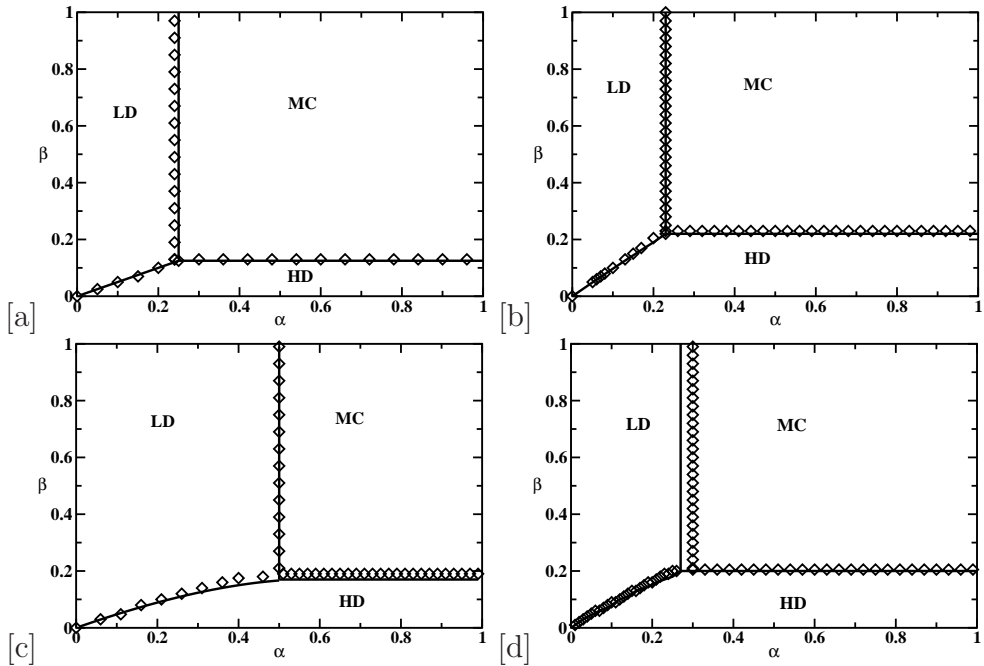

\centering [a]\includegraphics[scale=0.25,clip=true]{Fig5a.eps}
           [b]\includegraphics[scale=0.25,clip=true]{Fig5b.eps}
           [c]\includegraphics[scale=0.25,clip=true]{Fig5c.eps}
           [d]\includegraphics[scale=0.25,clip=true]{Fig5d.eps}
\caption{Phase diagrams of a two-channel system that couples TASEP and SSEP: a)  $w_1=w_2=1$; b) $w_1=w_2=1/3$; c)  $w_1=1$ and $w_2=0$; and d)  $w_1=0.4$ and  $w_2=0.25$. Symbols correspond to Monte Carlo simulations, while  theoretical predictions are represented by lines.}
\end{figure}

\section{Monte-Carlo simulations and discussions}

We presented two theoretical approaches to investigate parallel coupling of TASEP and SSEP. In the first approach, the mapping of two-channel systems into  effective single-lane exclusion models with known stationary properties has been utilized for strong couplings. This provides an exact description of all dynamics at large times. However, when the strength of the couplings between the channels was not large, we utilized the approximate mean-field method that neglects horizontal correlations in the system.  In order to check the validity of the approximate method and to examine theoretical predictions we performed extensive Monte Carlo computer simulations. 

The obtained theoretical results are valid only in thermodynamic limit, $L \rightarrow \infty$. In our simulations we used $L=1000$ for each channel, although in several cases we checked our computations also for lattices with $L=500$. It was found that computed dynamic properties do not depend on the size of the lattices, suggesting that finite-size effects are negligible in our simulations. The density profiles and the particle currents were calculated by averaging over trajectories that had between $2 \times 10^{6}$ and $10^{8}$ Monte Carlo steps. To ensure that the system reached the stationary state, first 5$\%$ of the total number of steps were ignored in averaging procedures. Phase transitions between phases were determined by observing the abrupt changes in the density profiles for transitions between HD and LD phases. For boundaries between HD or LD and MC phases the transition points  were determined  by observing the saturation of the particle current. These procedures ensure that phase border lines are determined with precision within 0.01 units of $\alpha$ and $\beta$.

Density profiles for different symmetric and asymmetric couplings between TASEP and SSEP channels  are presented in figures 2, 3 and 4. In all situations excellent agreement between Monte Carlo computer simulations and theoretical predictions is observed for bulk densities. Different behavior is found for symmetric and asymmetric couplings between the lattices. Equal vertical transitions rates make the properties of both channels almost the same, with slight differences near boundaries, especially for LD (figures 2a and 2b) and for MC phases (figures 4a and 4b). Increasing the strength of the symmetric coupling puts more particle in the system, and even for the LD phase the bulk densities are larger than 1/2: see figures 2a and 4a. Surprising results are found for the HD phase where bulk densities are  functions of only the exit rate $\beta$ and they are independent of the strength of the coupling. For asymmetric coupling densities in the upper and lower channels differ significantly.  When the vertical transition rate from the TASEP to SSEP is larger it leads to larger densities in the symmetric lattice, and in the strong coupling limit ($w_{1}=1$ and $w_{2}=0$) it even fills the second channel completely in all phases. For intermediate asymmetric couplings both channels behave qualitatively similarly.

Phase diagrams for a two-channel systems that couples TASEP and SSEP  are illustrated in figure 5. In all cases the system can exist in one of three stationary phases: the entrance-dominated low-density phase, the exit-dominated high-density phase and the maximal-current phase specified by bulk dynamics.  Comparison between computer simulations and theoretical calculations suggests that our theoretical method quantitatively correct in description of stationary properties of this system.  However, there are several small deviations between theoretical predictions and Monte Carlo results,  especially for LD/MC phase transitions line for intermediate couplings (see figure 5d), indicating that correlations inside the lattice are important for some ranges of parameters. It can be seen that symmetric couplings between the channels decrease the phase volume for the high-density phases, while asymmetric couplings have the same effect on the LD phases. In the case of unequal  vertical transition rates and for intermediate symmetric couplings the boundaries between the LD and HD phases are not linear, as found for strong symmetric couplings, but rather slightly curved.

Inter-channel particle transitions influence the overall current through the system as shown in figure 6 for the MC phases. For any symmetric couplings the particle fluxes through the system will go down, as was found before in the case of two-channel TASEP systems \cite{pronina04}.  However, breaking the symmetry in the vertical transition rates actually leads to increase in the particle current. As illustrated in figure 6, for $w_{1}=1$ lowering the transition rate $w_{2}$ from 1 to zero increase the particle current in two times. Similar behavior is observed for other stationary phases. Since the two-channel system that couples asymmetric and symmetric exclusion processes might be relevant for understanding transport of motor proteins along protein filaments \cite{klumpp03}, we can argue that this observations might be important for understanding motor protein's dynamics. One can suggests that the flux of molecular motors can be controlled  by modifying the association and dissociation rates to protein filaments, e.g., via changing the ionic strength, temperature or viscosity.

\begin{figure}[h]
\centering \includegraphics[scale=0.3,clip=true]{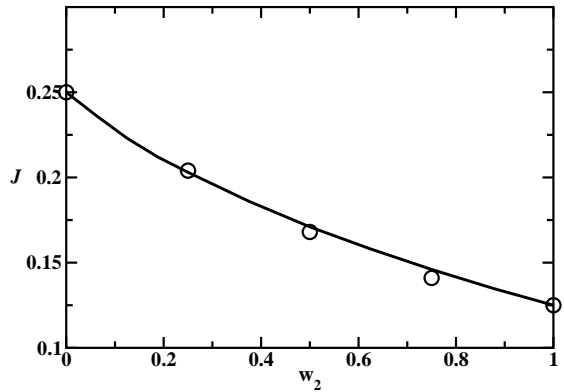} \caption{Particle current through a two-channel system that couples TASEP and SSEP as a function of the vertical transition rate $w_{2}$ for the fixed vertical transition rate $w_{1}=1$ and for $\alpha=\beta=0.75$. Symbols are from Monte Carlo simulations and lines are the result of theoretical calculations.}
\end{figure}

\section{Summary and conclusions}

We investigated the two-channel system that couples asymmetric and symmetric exclusion processes for different inter-channel transition rates in steady-state regime. In the limit of strong symmetric coupling, $w_{1}=w_{2}=1$, the exact description of particle dynamics is achieved by mapping the two-channel system, consisting of totally asymmetric and symmetric exclusion processes, into an effective one-channel  partially asymmetric exclusion process with known stationary properties. Exact solutions are also obtained in the limit of strong asymmetric inter-channel rates ($w_{1}=1$ and $w_{2}=0$, or $w_{1}=0$ and $w_{2}=1$). In this case the two-channel system  mapped into a single-lane totally asymmetric process with explicit description of all dynamic properties.

The two-channel system with coupled TASEP and SSEP for intermediate vertical transition rates has been analyzed via an approximate theoretical approach. In this method the dynamics of vertical inter-channel transitions is fully accounted, while the correlations along the horizontal lattices are neglected. The vertical-cluster mean-field approach allowed us to calculate analytically or numerically exactly particle currents, bulk densities and phase diagrams. The predictions of the approximate method are in excellent agreement with extensive Monte Carlo computer simulations. Theoretical calculations and computer simulations indicate that the strength and symmetry in the vertical transition rates have a strong effect on the overall particle dynamics. Equal inter-channel transition rates symmetrize the particle properties in both channels. Symmetric couplings  also lower the particle fluxes and increase the bulk densities. Asymmetric inter-channel transitions generally lead to similar qualitative properties in both channels, although with different values of bulk densities and currents. Asymmetric couplings also increase the transport capability of the system.

We discussed the relevance of the results of this investigation for understanding mechanisms of motor protein's motion along the protein filaments. It can be concluded from our theoretical analysis that dynamics of motor proteins can be controlled and modified by changing the association and dissociation rates. It will be interesting to test our predictions in experimental studies.

\section{Acknowledgments}The authors acknowledge the support from the Welch Foundation (under grant no. C-1559) and from the US  National Science Foundation through the grant NIRT ECCS-0708765.

\end{document}